\documentclass[fleqn,twoside]{article}
\usepackage{espcrc2}

\newcommand{\ltsima}{$\; \buildrel < \over \sim \;$}
\newcommand{\simlt}{\lower.5ex\hbox{\ltsima}}
\newcommand{\gtsima}{$\; \buildrel > \over \sim \;$}
\newcommand{\simgt}{\lower.5ex\hbox{\gtsima}}

\def\lesssim{\mathrel{\hbox{\rlap{\hbox{\lower4pt\hbox{$\sim$}}}\hbox{$<$}}}}
\def\gtrsim{\mathrel{\hbox{\rlap{\hbox{\lower4pt\hbox{$\sim$}}}\hbox{$>$}}}}

\def\ab1450{$AB_{1450(1+z)}$}

\def\xray{\hbox{X-ray}}

\def\today{\ifcase\month\or January\or February\or March\or April\or May\or
      June\or July\or August\or September\or October\or November\or December\fi
      \space\number\day, \number\year}
\def\asca{{\it ASCA\/}}
\def\chandra{{\it Chandra\/}}

\def\heao1{{\it HEAO-1\/}}

\def\rosat{{\it ROSAT\/}}

\def\sax{{\it BeppoSAX\/}}

\def\xmm{{XMM-{\it Newton\/}}}

\usepackage{graphicx}
\usepackage[figuresright]{rotating}

\newcommand{\AmS}{{\protect\the\textfont2
  A\kern-.1667em\lower.5ex\hbox{M}\kern-.125emS}}

\title{Restless Quasar Activity: From \sax\ to \chandra\ and \xmm}

\author{C. Vignali,\address[INAF-OAB]{INAF-Osservatorio Astronomico di Bologna, 
        Via Ranzani, 1, 40127 Bologna, Italy}
        \thanks{Support from the Italian Space Agency under contract ASI I/R/073/01 is acknowledged.} 
        A. Comastri\addressmark[INAF-OAB]
	and
        W.N. Brandt\address[PSU]{Department of Astronomy and Astrophysics, The Pennsylvania State University, 
525 Davey Lab, University Park, PA 16802, USA}
        }

\begin{document}

\begin{abstract}
We briefly review some of the progress made in the last decade in the study of the \xray\ properties of 
the quasar population from the luminous, local objects observed by \sax\ to 
the large, rapidly increasing population of $z>4$ quasars 
detected by \chandra\ and \xmm\ in recent years. 
\vspace{1pc}
\end{abstract}

\maketitle

\section{Introduction}
Over the last decade \asca\ and \sax\ have significantly improved our knowledge 
of the \xray\ properties of Active Galactic Nuclei (AGNs), especially 
at low redshifts, thanks to their broad-band coverage and relatively large effective areas above 2~keV. 
Moreover, the unique \xray\ coverage above $\approx$~10~keV (up to \hbox{$\approx$~100--200~keV}) 
provided by the Phoswich Detector System (PDS) onboard \sax\ has allowed proper definition of 
intrinsic \xray\ continuum shapes of local Seyfert galaxies (e.g., \cite{m01,r02,dc03,mal03}; 
see also \cite{pd03}, this Volume) and quasars (e.g., \cite{mineo00}). 

In Section~2 we will briefly recall the most important \sax\ results obtained for a peculiar AGN: the 
luminous, nearby radio-quiet quasar (RQQ) PDS~456. 

Although a few \xray\ spectral studies of quasars at 
$z\approx$~2--3 have been carried out with \asca\ (e.g., \cite{cappi97,v99}) and 
\sax\ (e.g., \cite{elvis00}), 
the properties of luminous RQQs at higher redshifts ($z>4$) were poorly known before the 
launches of the current generation of \xray\ telescopes, \chandra\ and \xmm. 
In Section~3 we will discuss the large improvements that have occurred in this field in the last few years.

\section{One ``intriguing'' \sax\ observation: \\ 
Warm and cold absorption in the luminous, nearby RQQ PDS~456}
Unfortunately, high-luminosity quasars are usually found at 
relatively high redshifts, thus appearing rather weak and difficult 
to study in X-rays before the launches of \chandra\ and \xmm. 
%
\begin{figure*}
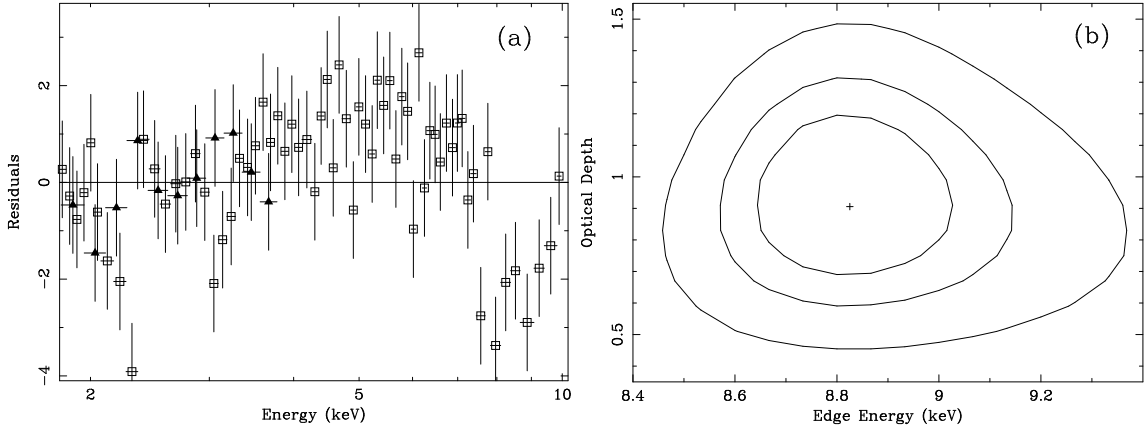

\includegraphics[height=\columnwidth,angle=-90]{vignali.fig1a.ps}
\includegraphics[height=\columnwidth,angle=-90]{vignali.fig1b.ps}
\vglue-0.6cm
\caption{(a) Data-to-model residuals for a single power-law fit to the 
LECS (triangles) and MECS (squares) data of PDS~456. 
(b) 68, 90, and 99\% confidence regions for the rest-frame edge energy versus 
the optical depth derived from the \sax\ spectrum of PDS~456 assuming model (B) in \cite{v00}.}
\label{fig:fig1}
\end{figure*}
%
In this context, PDS~456, at $z=0.184$ and with 
\hbox{$L_{\rm bol}$$\simeq$10$^{47}$~erg~s$^{-1}$}, 
can be considered an exceptional case, 
thus allowing an accurate modeling of the \xray\ continuum and 
reprocessing features with both \sax\ and \asca\ (\cite{v00}) and, recently, 
with \xmm\ (\cite{r03}). 
The \xray\ spectrum of PDS~456 is characterized by 
a prominent ionized Fe\ K~edge [clearly visible in the data-to-model 
residuals shown in Fig.~1, panel (a), when a single power-law fit is adopted]; 
the edge corresponds to Fe~{\sc xxiv-xxvi} at $\approx$~8.8~keV 
[see Fig.~1, panel (b)]. 
The lack of iron emission lines suggests that the ionized edge 
is due to matter along the line-of-sight rather than reflection from a 
highly ionized accretion disk. 
%
%
Indeed, the hard \xray\ continuum is due to transmission through a very ionized medium, 
best fitted by a column density \hbox{$N_{\rm H_{\rm warm}}\approx4.5\times10^{24}$~cm$^{-2}$}, 
coupled with absorption by cold matter having 
\hbox{$N_{\rm H_{\rm cold}}\approx2.7\times10^{22}$~cm$^{-2}$} (\cite{v00}; see Fig.~2 for the 
best-fit spectrum). 
%
%

The \xray\ properties of PDS~456 appear quite different from those of the majority of 
the local Palomar-Green quasars (e.g., \cite{mineo00,g00}); 
its photon index is rather flat ($\Gamma=1.4-1.6$; see \cite{v00}) and 
the absorber ionization parameter $U=\frac{n_{\rm phot}}{n_{\rm e}}$, defined as the ratio of the 
ionizing photon density at the surface of the cloud 
to the electron density of the gas, is extremely high ($\approx$~7900; see \cite{v00}).
%
%
%
This overall \xray\ picture for PDS~456 has been confirmed recently by \xmm, whose 
spectral resolution has allowed three high-ionization iron~K edges (\cite{r03}) 
to be distinguished instead of the one observed by \sax\ 
(likely due to the different spectral resolution and effective area). 
\xmm\ has also discovered an extreme gas outflow velocity of $\approx$~50,000~km~s$^{-1}$ (\cite{r03}), 
thus supporting the idea that the ionized matter is close to the active nucleus of PDS~456. 
Furthermore, the source showed repeated \xray\ flaring episodes, with an \xray\ flux doubling time 
of $\approx$~30~ks and a total energy output of the flaring events as high as 10$^{51}$~erg 
\cite{reeves02}. 
This extreme \xray\ variability and the presumably high accretion rate 
make this source more similar to the Narrow-Line Seyfert~1 galaxies. 
%
High-resolution \xray\ spectroscopy of such objects can 
provide further details on the accretion mechanisms responsible 
for the \xray\ emission in high-luminosity objects (e.g., \cite{r03,pounds03}). 
%
\begin{figure}[!h]
\includegraphics[height=\columnwidth,angle=-90]{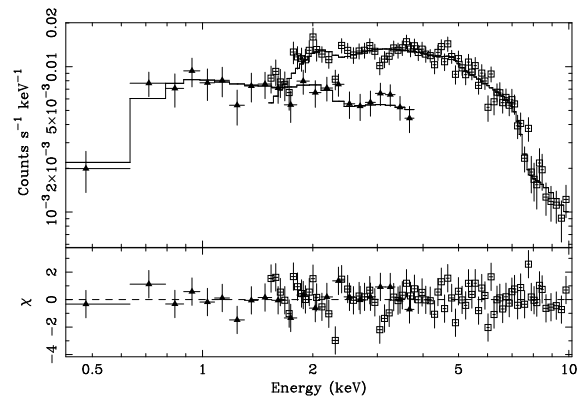}
\vglue-0.6cm
\caption{\sax\ LECS (triangles) and MECS (squares) 0.5--10~keV spectrum of PDS~456. 
The data-to-model residuals are shown in the bottom panel in units of $\sigma$.}
\label{fig:fig2}
\end{figure}

\section{The realm of the ancient quasars}
The last few years evidenced an increasing interest 
in the study of $z\simgt4$ AGNs (mainly quasars), including in the \xray\ band. 
Prior to 2000 there were only six quasars at $z>4$ with \xray\ detections. 
The number of \xray\ detected quasars at $z>4$ doubled when the 
first systematic \xray\ study of these objects was carried out using 
archival \rosat\ data (\cite{k00}). 
Since then, the progress made in this field has been substantial. 
This has been possible due to the availability of increasing numbers of 
$z>4$ quasars from ground-based optical surveys 
(e.g., the Sloan Digital Sky Survey -- SDSS, \cite{y00}, 
the Digital Palomar Sky Survey -- PSS, \cite{d98}, and 
the Automatic Plate Measuring facility survey -- APM, \cite{i91}),\footnote{See 
http://www.astro.caltech.edu/$\sim$george/z4.qsos for a listing of high-redshift quasars.}
and the excellent capabilities of \chandra\ and \xmm\ for detecting faint sources. 

To define the basic individual \xray\ properties (i.e., \xray\ fluxes, luminosities, and 
optical-to-\xray\ spectral indices) of $z>4$ quasars, 
we started a program to observe with \chandra\ and \xmm\ both the 
optically brightest \hbox{$z\approx$~4--4.6} PSS/APM quasars and 
the higher redshift, optically fainter SDSS quasars (\cite{b01,v01,b02,v03a,v03b}; 
see also the recent review by \cite{b03}). 
Since the pioneering work with \rosat\ (\cite{k00}), 
the number of AGNs with \xray\ detections has therefore increased significantly 
to more than 80 (see 
Fig.~3),\footnote{See http://www.astro.psu.edu/users/niel/papers/highz-xray-detected.dat 
for a regularly updated listing of \xray\ detections and sensitive upper limits at $z>4$.}
in the redshift range $z\approx$~4--6.3 (also see 
\hbox{\cite{sch98,v02,sil02,be03,cas03}}).
%
\begin{figure}[!t]
\includegraphics[height=\columnwidth,angle=-90]{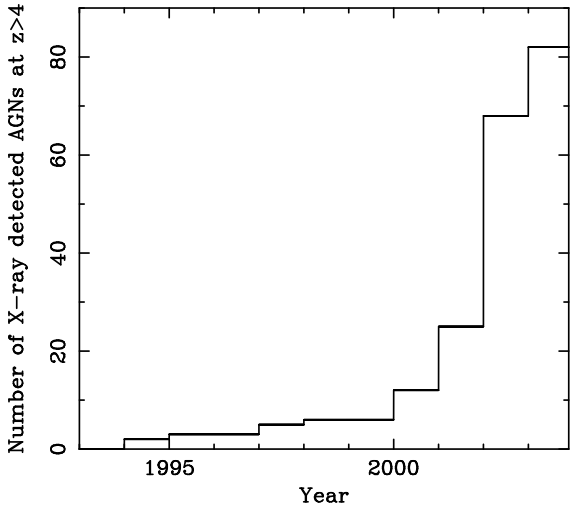}
\vglue-0.6cm
\caption{Cumulative number of $z>4$ \xray\ detected AGNs as a function of year.}
\label{fig:fig3}
\end{figure}
%
\begin{figure}[!th]
\includegraphics[height=\columnwidth,angle=0]{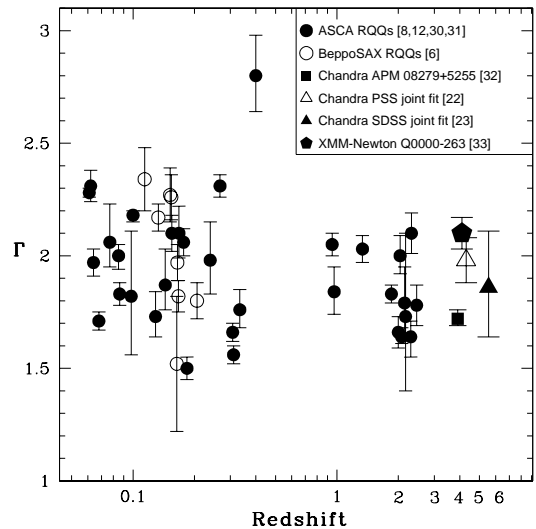}
\vglue-0.6cm
\caption{Plot of photon index versus redshift for optically selected RQQs. Different 
symbols indicate different samples (as shown in the key; also see \cite{v03b} for details). 
No clear indication of \xray\ continuum evolution with redshift is seen.}
\label{fig:fig4}
\end{figure}
%
Due to the extremely low background in typical \chandra\ snapshot ($\approx$~4--10~ks) 
observations, it has also been possible to derive average spectral constraints 
for subsamples of high-redshift quasars 
using joint spectral fitting with \hbox{$\approx$~120--340} \xray\ counts (\cite{v03a,v03b}). 
At $z>4$, optically selected RQQs have a photon index of 
\hbox{$\Gamma\approx$~1.8--2.0}, 
similar to the results found at low and intermediate redshifts (e.g., \cite{g00}). 
Furthermore, no spectral evolution of the \xray\ continuum shape over cosmic time has been 
found (see \cite{v03b} and Fig.~4). At high redshift, this result has been supported recently by 
direct \xray\ spectroscopy of QSO~0000$-$263 at $z=4.10$ with \xmm\ (\cite{fb03}). 

%

\end{document}